\pdfoutput=1
\documentclass[pra, showpacs, twocolumn, floatfix]{revtex4}
\usepackage{latexsym}
\usepackage{amsmath}
\usepackage{amsfonts}
\usepackage{graphicx}
\usepackage{exscale}
\usepackage{amssymb}
\usepackage{mathrsfs}

\newcommand{\mf}{\mathbf}
\newcommand{\imag}{{\rm i}}
\newcommand*{\D}{\mathrm{d}}
\newcommand{\U}[1]{\,{\rm{#1}}}
\newcommand{\euler}{\textrm{e}}
\newcommand{\au}{\U{a.u.}}
\newcommand{\eV}{\U{eV}}

\newcommand{\I}[1]{_{\mathrm{#1}}}

\begin{document}
\title{Macroscopic aspects of relativistic x-ray assisted\\ high-order harmonic generation}
\author{Markus C. Kohler}
\author{Karen Z. Hatsagortsyan}
\email{k.hatsagortsyan@mpi-k.de}
\affiliation{ Max-Planck-Institut f\"ur Kernphysik, 
Saupfercheckweg 1, 69117 Heidelberg, Germany}

\date{\today}

\begin{abstract}
A theoretical model is developed describing high-order harmonic generation (HHG) from  a gas of multiply-charged ions driven by a laser field of relativistic  intensity. Macroscopic propagation of harmonics is investigated in a relativistic HHG setup where the relativistic drift is suppressed by means of x-ray field assistance of the driving laser field. The possibility of phase-matched emission of the harmonics is shown. The laser field geometry is optimized to maximize the HHG yield with the corresponding phase-matching schemes. Crucial issues determining the macroscopic HHG yield are discussed in detail.

\end{abstract}


\maketitle

\section{Introduction}

High-order harmonic generation (HHG) is a reliable source of coherent soft x-ray radiation in the nonrelativistic regime. With the state-of-the-art technology, coherent x-rays of up to $\sim3$~keV~\cite{Seres:XA-06} of photon energy can be generated. 
The most favorable conversion efficiency for nonrelativistic multi-keV harmonics is anticipated with mid-infrared driving laser fields at high gas pressures~\cite{Popmintchev,Kapteyn,Murnane}.
A further increase of the photon energy can in principle be achieved by
increasing the laser intensity. However, the applicable laser intensity is
limited in two ways~\cite{Piazza:Re-12}. First, the relativistic electron drift prevents recollision
and results in a dramatic suppression of the HHG efficiency. And second, the
strong field causes rapid ionization of the medium leading to a large free
electron dispersion and along with a significant phase-mismatch.  The electron
recollision  is suppressed when the drift distance becomes larger than the
electron wave packet size at the moment of recollision~\cite{Walker}. This
happens when the laser intensity exceeds $10^{16}-10^{17}$ W/cm$^2$ for infrared
(IR) wavelengths. The ponderomotive potential of the laser field in this case
amounts to $U_p\approx 3$ keV and the achievable cutoff frequency for HHG
to $\omega_{c}\approx 10$ keV. This indicates the limit of nonrelativistic HHG.

Various methods to counteract the relativistic drift have been proposed. To suppress the drift, different laser field geometries, in some cases with an additional field, can be applied~\cite{Lin:HH-06,KYLSTRA:BS-00,TARANUKHIN:RH-00,VERSCHL:SW-07,MILOSEVIC:DA-04,LIU:LG-09,CHIRILA:ND-02,VERSCHL:RC-07,VERSCHL:RF-07,KLAIBER:TA-06,KLAIBER:LO-07,KLAIBER:CH-08,HATSAGORTSYAN:UFO-08,VERSCHL:LR-08}. Highly charged ions moving relativistically~\cite{MOCKEN:BA-04,CHIRILA:IS-04} or a gas of positronium atoms~\cite{HENRICH:PI-04,HATSAGORTSYAN:ML-06} can also be employed for this purpose. However, all these efforts have only addressed the drift suppression problem for the emission from a single atom rather than coherent emission from a macroscopic gas target where phase-matching becomes crucial. For the first time both 
problems of relativistic HHG, namely, the relativistic drift and the phase-matching, have been solved at the same time in~\cite{KOHLER:HX-11} where the macroscopic yield of HHG has been calculated in the setup consisting of two counterpropagating attosecond pulse trains~\cite{HATSAGORTSYAN:UFO-08}. In this setup, the relativistic drift caused by the ionizing laser pulse is reverted by the counter-propagating pulse inducing recombination. It appears that, specific to this setup, an additional harmonic phase exists which depends on the time delay between the driving pulse trains. We have shown in~\cite{KOHLER:HX-11} that this additional phase of the emitted harmonics can be tuned to compensate the phase mismatch caused by the free electron background. However, the setup of counter-propagating pulses is rather challenging, for instance, the requirement for a small pulse distortion imposes a rather strong restriction on the medium length. Additionally, a precise modulation of the laser intensity along the propagation direction is required.

Another appealing scheme for relativistic HHG exists based on XUV assistance~\cite{KLAIBER:CH-08} which seems experimentally less demanding than the scheme with counter-propagating attosecond pulses. 
The usefulness of XUV light assisting a strong laser field has been demonstrated in the nonrelativistic regime for various purposes. 
It has been used to enhance HHG by many orders of magnitude compared with the case via a fundamental laser pulse alone~\cite{ISHIKAWA:PI-03,TAKAHASHI:DE-07}.
When the XUV field has the form of an attosecond pulse train  a single quantum path can be selected to contribute to HHG  and in this way allowing to manipulate the time-frequency properties of harmonics as well as to enhance a selected bandwidth of harmonics~\cite{SCHAFER:PC-04,GAARDE:DE-05,FARIA:HO-07}. Tuning the XUV field to a resonance between a core and valence state can lead to the emergence of a second plateau that is shifted to higher energies by the former resonance energy with respect to the first plateau~\cite{Buth:HO-up}.

In the relativistic regime the XUV/x-ray assistance can be employed to overcome the relativistic drift motion~\cite{KLAIBER:CH-08}. Thereby, the XUV frequency requires to exceed the ionization energy to 
liberate the electron with a single photon and to deliver a significant initial momentum to the freed electron. This way, the electron can obtain sufficient momentum in the direction opposite to the laser propagation direction to  compensate for subsequent drift motion and return to the atomic core, recombine and emit harmonics after the excursion in the relativistically strong laser field.  
The medium  is a gas of multiply charged ions with an ionization energy large enough to withstand the strong optical laser field.
How much the XUV assisted setup   for relativistic HHG favors phase-matching needs investigation.
Formalisms describing macroscopic effects due to ionization and phase matching in HHG 
are restricted to the non-relativistic regime, for a recent review see, e.g.,~\cite{GAARDE:MA-08}.

In this paper, we investigate the feasibility of phase-matched emission and the macroscopic yield of harmonics  in the relativistic regime of the x-ray assisted HHG setup in a strong IR laser field.
Generally, the efficiency of HHG is rather small even in the nonrelativistic regime due to the wave packet spreading. In the relativistic regime, the single-atom HHG emission rate continues to decrease even when the relativistic drift is compensated~\cite{KOHLER:HX-11}. Thus, a large phase-matching volume is crucial in order to achieve a significant HHG yield. Furthermore, the large ponderomotive potential is likely to result in rapid phase changes if ions emit under different conditions. For generating relativistic harmonics  both challenges have to be met: circumventing the drift and having the setup stable against phase changes. Our presented setup overcomes both issues and renders a measurable HHG yield in the relativistic regime possible.

The structure of the paper is the following. In Sec.~\ref{sec:relHHG_macr}, the theory of macroscopic HHG is presented applicable for any field geometry in the relativistic regime. In Sec.~\ref{sec:relxray}, the developed theory is applied to calculate the macroscopic HHG yield for the setup  of x-ray assisted relativistic HHG. Our conclusion is presented in Sec.~\ref{sec:conclusion}.

\section{Macroscopic model for relativistic HHG}\label{sec:relHHG_macr}

\subsection{Macroscopic HHG yield}\label{sec:relHHG_macr1}

\begin{figure}[b]
\begin{center}
 \includegraphics[width=0.48\textwidth]{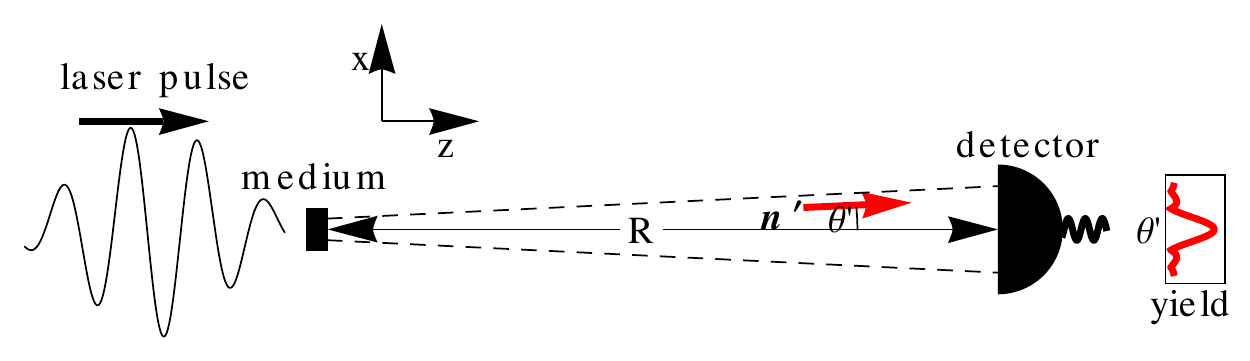}
\caption{(color online). Geometry of the medium and detector including the definitions of the coordinates. The dashed lines denote the divergence angle of the harmonic radiation. The box on the right schematically shows the measured angular distribution.} \label{HHG_geometry}
\end{center}
\end{figure}

In this section, our model is presented for the calculation of 
the harmonic spectrum from a macroscopic gas target suitable for relativistic laser intensities. In the non-relativistic regime, the standard approach for the calculation of the macroscopic HHG response incorporates  the single-atom contribution via the time-dependent dipole moment~\cite{HUILLIER:TA-91,PRIORI:PM-00,GAARDE:MA-08}. However, employing the dipole moment for the radiation response assumes that an emitted harmonic wavelength is much longer than the spatial extensions of the emitter. This approach fails for sufficiently small wavelengths because then the retardation between different points of the emitting wave packet becomes important~\cite{HERNANDEZ:HP-91,MOCKEN:RS-05}. Our approach uses the complete current density distribution of each atom  rather than the dipole moment. Retardation between different emission points within the distribution is taken into account by a phase factor.
The link between the microscopic (atomic) current density $\mf{j}$ and the macroscopically emitted harmonic electric field $\mf{E}\I{H}$ is obtained from Maxwell's equations similar to the non-relativistic approaches~\cite{HUILLIER:TA-91,PRIORI:PM-00,GAARDE:MA-08}. The Fourier component of the emitted harmonic electric field from a gas target  is given by~\cite{JACKSON:CE-98}
\begin{equation}\label{eq:EH_general}
 \tilde{\mf{E}}\I{H}(\mf{x}',\omega\I{H})=\imag \frac{\omega\I{H}}{c^2} \int\D^3 \textbf{x}
\frac{\mf{n}'\times(\tilde{\mf{j}}(\mf{x},\omega\I{H})\times \mf{n}')}{R}
\euler^{\imag k\I{H} R}\, ,
\end{equation}
where $\tilde{\mf{j}}(\mf{x},\omega\I{H})$ is the Fourier component of the current density,
$\omega\I{H}$ is the frequency of the emitted harmonic light, $\mf{k}\I{H}=k\I{H}\mf{n}'$  the wave vector, $R=\vert \mf{x}-\mf{x'}\vert$ the distance between the emission and observation point, $\mf{x'}$ the coordinate of the observation point  and 
$\mf{n}'=\mf{x}'/|\mf{x}'|$ the unit vector in the observation direction (see Fig.~\ref{HHG_geometry}).
Absorption of the harmonic photons is neglected~\cite{PRIORI:PM-00} because their energy is much higher than the largest atomic transition energy. The current density $\mf j$ is exclusively determined by the HHG process. For the evaluation of Eq.~\eqref{eq:EH_general}, we restrict ourselves 
to the far-field zone which is sufficient for calculating the overall HHG photon yield. The far-field zone is determined by the conditions that the distances from the emitters to the observation point are larger than the wavelength of the emitted radiation as well as the size $a$ of the emitting region ($k\I{H} R\gg1$ and $R\gg a$).  We  thus can expand Eq.~\eqref{eq:EH_general} over a small parameter $a/R$ using 
$R=\vert \mf{x}-\mf{x'} \vert \simeq |\mf{x}'| - \mf{x}\cdot \mf{n}'$~\cite{JACKSON:CE-98}. When inserting this expression into Eq.~\eqref{eq:EH_general}, the exponential function splits up into two parts. One term contains $\textbf{x}'$ and is thus a general phase factor depending on the constant observation point that can be separated: $\tilde{\mf{E}}\I{H}(\mf{x}',\omega\I{H})=\euler^{\imag k\I{H} |\textbf{x}'|}\tilde{\mf{E}}\I{H,0}(\mf{n}',\omega\I{H})$. In the following we consider $\tilde{\mf{E}}\I{H,0}(\mf{n}',\omega\I{H})$ only and find
\begin{equation}
  \tilde{\mf{E}}\I{H,0}(\mf{n}',\omega\I{H})=\imag \frac{\omega\I{H}}{c^2R}\int\D^3 \mf{x}\,  \mf{n}'\times(\tilde{\mf{j}}(\mf{x},\omega\I{H})\times \mf{n}') \euler^{-\imag \mf{k}\I{H}\cdot \mf{x}} \label{E01}\, .
\end{equation}
 The total current density distribution consists of a sum of the current densities  of the single atoms $\mf{j}\I{a}(\mf{x}\I{a},\mf{x},t)$ with positions $\mf{x}\I{a}$:
\begin{equation}
 \tilde{\mf{j}}(\mf{x},\omega\I{H})=\int \D^3\textbf{x}\I{a} \rho(\textbf{x}\I{a})\int \D t\, \mf{j}\I{a}(\mf{x}\I{a},\mf{x},t) \euler^{\imag \omega\I{H} t} \label{curint}
\end{equation}
where $\rho(\textbf{x}\I{a})$ is the atomic number density. 
Inserting  Eq.~\eqref{curint} into~Eq.~\eqref{E01} yields the final expression for the macroscopically emitted harmonic field
\begin{eqnarray}
  \tilde{\mf{E}}\I{H,0}&&\!\!\!\!\!\!(\mf{n}', \omega\I{H})\nonumber\\
&&\!\!\!\!\!\!=\imag \frac{\omega\I{H}}{Rc^2} \int \D^3\textbf{x}\I{a}  \rho(\textbf{x}\I{a}) \int \D^3 \textbf{x} \nonumber\\ &&\int \D t\,\mf{n}'\times(\mf{j}\I{a}(\mf{x}\I{a},\mf{x},t)\times \mf{n}') \euler^{-\imag k\I{H} \mf{x}\cdot{\mf{n}'}+\imag \omega\I{H} t} \label{E02}\nonumber\\
&&\!\!\!\!\!\!= \imag \frac{\omega\I{H}}{Rc^2} \int \D^3\mf{x}\I{a}\, \rho(\textbf{x}\I{a})\,  \mf{n}'\times(\tilde{\mf{j}}\I{a}(\mf{x}\I{a},\mf{n}',\omega\I{H})\times \mf{n}')\label{phase_match_integral}\, ,
\end{eqnarray}
where $\tilde{\mf{j}}\I{a}(\mf{x}\I{a},\mf{n}',\omega\I{H})\equiv \int \D^3 \textbf{x}\int\D t\mf{j}\I{a}(\mf{x}\I{a},\mf{x},t)\exp\{-\imag k\I{H} \mf{x}\cdot{\mf{n}'}+\imag \omega\I{H} t\}$.
Note that the combination of outer products  in Eq.~\eqref{phase_match_integral} prevents mathematically from emission in the direction of the current density vector.
However, due to phase-matching, the macroscopic emission is in many cases mainly along the propagation direction of the laser $\mf{n}'\approx\hat{\mf z}$, (see Fig.~\ref{HHG_geometry}). When the laser is linearly polarized in $\hat{\mf x}$ direction, the current density vector is parallel  to the $\hat{\mf x}-\hat{\mf z}$ plane. Thus, we can approximate
$\mf{n}'\times(\tilde{\mf{j}}\times\mf{n}')=\tilde{\mf{j}}-(\mf{n}'\cdot\tilde{\mf{j}})\mf{n}'\approx \tilde{j\I{x}} \hat{\mf x}$
and can restrict ourselves to the $\hat{\mf x}$ component
\begin{eqnarray}
  \tilde{E}\I{H,0,x}(\mf{n}', \omega\I{H})
= \imag \frac{\omega\I{H}}{Rc^2} \int \D^3\textbf{x}\I{a}\,\rho(\textbf{x}\I{a})  \tilde{j}\I{a,x}(\mf{x}\I{a},\mf{n}',\omega\I{H})\label{phase_match_integralEx}
\end{eqnarray}
to describe emission in this case.

The overall emitted energy can be obtained via integrating the Poynting vector $\mf{S}(\mf{r},t)=\frac{c}{4\pi}\mf{E}^2\I{H}(\mf x', t)\mf{n}'$ overall emission directions  in the far field
\begin{equation}
 W=\frac{c R^2}{4\pi}\int \D t \int  \D \Omega\, \mf{E}\I{H}^2(\mf{x},t) \label{W_integral0}\, .
\end{equation}
By inserting 
\begin{equation}
 \mf{E}\I{H}(\mf{x}',t)=\frac{1}{2\pi}\int_{-\infty}^{\infty} \D \omega\I{H} \euler^{-\imag \omega\I{H} t}\tilde{\mf{E}}\I{H}(\mf{n}',\omega\I{H})\, 
\end{equation}
into ~Eq.~\eqref{W_integral0}, the energy can be calculated via an integration over the spectrum~\cite{LANDAU:CF-62}: 
\begin{equation}
 W=\frac{c R^2}{(2\pi)^2} \int \D \Omega \int_{0}^{\infty} \D \omega\I{H} \vert\tilde{\mf{E}}\I{H,0}(\mf{n}',\omega\I{H}) \vert^2.\label{W_integral}
\end{equation}
The emitted spectral photon number per solid angle from Eq.~\eqref{W_integral} is given by: 
\begin{equation}
\frac{\D N}{\D \omega\I{H} \D\Omega}=\frac{c R^2}{(2\pi)^2\omega\I{H}}  \vert\tilde{\mf{E}}\I{H,0}(\mf{n}',\omega\I{H}) \vert^2. \label{eq:spectr-phot-num}
\end{equation}

\subsection{Single-atom current density} \label{quantum_current}

In Sec.~\ref{sec:relHHG_macr1}, the macroscopic HHG yield has been calculated via classical electrodynamics. Since the expression Eq.~\eqref{phase_match_integralEx}  for the emission field relies on the current densities of a single atom in the gas, we continue to derive the  single-atom current density quantum mechanically in the relativistic regime via the Klein-Gordon equation. The Klein-Gordon current density of a particular atom at position $\mf{x}\I{a}$ in the laser field is given by~\cite{BJORKEN:RQ-64}
\begin{equation}\label{eq:KG-current}
\mf{j}\I{a}(\mf{x}\I{a},\mf{x},t)=\Bigl(\Psi^*_{\mf{x}\I{a}}(x)\hat{\mf{j}}\Psi_{\mf{x}\I{a}}(x)+\Psi_{\mf{x}\I{a}}(x)\hat{\mf{j}}^* \Psi^*_{\mf{x}\I{a}}(x)\Bigr) 
\end{equation}
where $ \hat{\mf{j}}=\hat{\mf{p}}+\mf{A\I{L}}(x)/c$ and  $\Psi_{\mf{x}\I{a}}(x)$ is the solution of the Klein-Gordon equation when the binding potential is centered around $\mf{x}\I{a}$. In the following, the time-space coordinate is $x=(ct,\mf{x})$, the wave four-vector of the laser field $k_L=(\omega\I{L}/c,\mf k\I{L})$, and the metric tensor $g_{\mu \nu}=\textrm{diag}(1,-1,-1,-1)$.

By a Fourier transformation of Eq.~\eqref{eq:KG-current} and partial integration, the spectral current is obtained:
\begin{eqnarray}
&&\tilde{\mf{j}}\I{a}(\mf{x}\I{a},\omega\I{H},\mf{n}') \nonumber\\ 
&=&\frac{1}{c}\int \D^4 x \,   \euler^{\imag k\I{H} x} \bigl(\Psi^*_{\mf{x}\I{a}}(x)\hat{\mf{j}}\Psi_{\mf{x}\I{a}}(x) +\Psi_{\mf{x}\I{a}}(x)\hat{\mf{j}}^* \Psi^*_{\mf{x}\I{a}}(x)\bigr) \nonumber \\
&=& \frac{2}{c}\int \D^4 x \,   \euler^{\imag k\I{H} x} \Psi_{\mf{x}\I{a}}^*(x)(\hat{\mf{j}}-\tfrac{1}{2} \mf{k}\I{H})\Psi_{\mf{x}\I{a}}(x) \label{eq:KG-jw}
\end{eqnarray}
We calculate the electron wave function in the field of the laser and the
ionic core by means of the strong-field approximation (SFA)~\cite{REISS:CK-90,REISS:RS-90} 
\begin{equation}\label{eq:SFA_WF}
 \Psi_{\mf{x}\I{a}}(x)=\phi(\mf{x}-\mf{x}\I{a},t)+\int d^4 x' G\I{L}(x,x')V\I{I}(x')\phi(\mf{x'}-\mf{x}\I{a},t')
\end{equation}
where $V\I{I}=2\imag(\textbf{A}\I{L}(\eta)/c)\nabla -\textbf{A}\I{L}^2(\eta)/c^2)$
is the term in the Hamiltonian describing the electron interaction with the
laser field with the vector-potential $\textbf{A}_L(\eta)$ and phase $\eta=
k\I{L}\cdot x$.
The Volkov propagator $G\I{L}(x,x')$ in a plane wave laser field is given by~\cite{MILOSEVIC:UH-00,MILOSEVIC:UH-02}
\begin{equation}\label{eq:relHH_Volk_prop}
  G\I{L}(x,x')=-\imag\,\theta(t-t')\int 
  \frac{c \,\D^3\mathbf{q} }{2\varepsilon_{\mathbf{q}} (2\pi)^3}
  \exp\big[-\imag\,S\I{L}(x,x')\big]\\
\end{equation} 
with the classical action of an electron in the laser field
\begin{equation}\label{eq:SL}
 S\I{L}(x,x')=q\cdot\left(x-x'\right)
+\int^{\eta}_{\eta'}\D\tilde{\eta}\left[\frac{\left(\mathbf{q}+ 
        \mathbf{A}_L(\tilde{\eta})/2c\right)\cdot \mathbf{A}\I{L}(\tilde{\eta})/c}
    {k\I{L}\cdot q}\right]\, ,
\end{equation}
the energy-momentum four-vector $q=(\varepsilon_{\mathbf{q}}/c,\mathbf{q})$, the
energy $\varepsilon_{\mathbf{q}}=\sqrt{c^2\mathbf{q}^2+c^4}$.
Here $\phi(x)=\frac{\phi_0(\mf{x})c}{\sqrt{2(c^2-I\I{p})}}\exp\{-\imag[(c^2-I\I{p})t+\mf{x}\cdot\mf{A}\I{L}/c]\}$, where  $\phi_0(\mf{x})$ is the nonrelativistic ground-state wave function. 
Inserting~Eq.~\eqref{eq:SFA_WF} into~Eq.~\eqref{eq:KG-jw} and applying the
usual assumptions~\cite{Lewenstein:HH-94,MILOSEVIC:UH-02} of neglecting  bound--bound and continuum--continuum transitions and the time--inverted process, we obtain
\begin{eqnarray}
&&\!\!\!\!\!\!\!\!\!\!\!\!\tilde{\mf{j}}\I{a}(\mf{x}\I{a},\omega\I{H},\mf{n}')\nonumber\\
&&\approx\frac{2}{c} \int \D^4 x\int \D^4 x'\euler^{\imag k\I{H}\cdot x}\phi^*(\mf{x}-\mf{x}\I{a},t)(\hat{\mf{j}}-\tfrac{1}{2} \mf{k}\I{H}) \nonumber\\
&&\,\,\,\,\times G\I{L}(x,x') V\I{I}(x')\phi(\mf{x'}-\mf{x}\I{a},t')\, .\label{j_expr_L}
\end{eqnarray}

\subsection{Electron wave function in a distorted plane wave laser field} \label{sec:eikonal}

The Volkov  propagator Eq.~\eqref{eq:relHH_Volk_prop} describes the evolution of the wave function of an electron in a plane wave laser field with vector potential $\mf{A}\I{L}(\eta)$.  However, in many practical situations this assumption on the laser field is not met, 
in particular, when a focused laser field or multiple laser beams are applied~\cite{PEATROSS:SZ-97,COHEN:GA-99,SERRAT:SE-10}, or when the dispersion distorts the laser pulse. For this reason, we need to find the electron wave function in an external laser field where the vector potential depends not only on phase $\eta$ but also on the position $z$ along propagation. 
The deviation of the laser field from the plane wave form is assumed to be a perturbation, so that the total vector potential reads  
\begin{equation}
\mf{A}(\eta,\mf{x})=\mf{A}\I{L}(\eta)+\mf{A}\I{P}(\eta,\mf{x}),
\label{A} 
\end{equation}
with $\vert\mf{A}\I{P}(\eta,\mf{x})\vert\ll \vert\mf{A}\I{L}(\eta)\vert$. We
find the solution of the Klein-Gordon equation in the field of Eq.~\eqref{A}
using the eikonal approximation, see e.g.~\cite{AVETISSIAN:GE-99,SMIRNOVA:EI-08},
in which the impact of the perturbation onto the wave function is taken into account by an expansion of the wave function phase. 
The electron wave function is described by the Klein-Gordon equation
\begin{equation}\label{eq:rel_KGeik}
 (\partial^\mu\partial_\mu +c^2)\Psi(x)=V\Psi(x)\, ,
\end{equation}
with $V=2\imag(\mf{A}(\eta)/c)\nabla-A^2(\eta)/c^2$. In order to solve Eq.~\eqref{eq:rel_KGeik}, the ansatz 
\begin{equation}\label{eq:rel_ansatzeik}
\Psi(x)=\frac{\sqrt{c}}{\sqrt{2(2\pi)^3\varepsilon_{\mf{p}}}}\exp\{-\imag[S\I{L}
(x,x\I{0})+S\I{P}(x,x\I{0})]\} \, .
\end{equation}
is employed. Here, $S\I{L}(x,x\I{0})$ is defined in Eq.~\eqref{eq:SL}, with an
arbitrary constant $x_0$, and, consequently, $\exp(-\imag S\I{L}(x,x\I{0}))$
solves the unperturbed equation \eqref{eq:rel_KGeik} with
$\mf{A}(\eta,z)=\mf{A}\I{L}(\eta)$.
Inserting the ansatz Eq.~\eqref{eq:rel_ansatzeik} into Eq.~\eqref{eq:rel_KGeik},
one finds
 \begin{eqnarray}
&&\frac{1}{c^2}[-\imag\partial_t^2 S\I{P} -2\partial_t S\I{L}\partial_t S\I{P} -(\partial_t S\I{P})^2 ]\nonumber\\
&&\;\;\;\;\;-[-\imag\nabla^2 S\I{P}-2\nabla S\I{L}\nabla S\I{P}-(\nabla S\I{P})^2]\nonumber\\
&&\;\;\;\;\;\;\;\;\;\;=2\frac{\mf{A}\I{L}}{c}\nabla S\I{P}+2\frac{\mf{A}\I{P}}{c}\nabla(S\I{L}+S\I{P})
\nonumber\\
&&\;\;\;\;\;\;\;\;\;\;\;\;\;\;\;\;-\frac{A\I{P}^2}{c^2}-2\frac{A\I{L} A\I{P}}{c^2}.
\end{eqnarray}
In the applied approximation, $|\nabla S\I{p}|\ll |\textbf{A}\I{P}/c|$ and
$|\partial_t S\I{P}|\ll |\textbf{A}_P|$, which allows to neglect the  $(\nabla
S\I{P})^2$ and $ (\partial_t S\I{P})^2$ terms. When additionally $\xi\equiv
E\I{L}/c\omega\I{L}\gg \omega\I{L}/c^2$, one has $|\nabla^2 S\I{p}|\ll
|(\textbf{A}_P/c)\nabla S\I{p}|$ and $|\partial_t^2 S\I{p}|\ll |\partial_t
S\I{L} \partial_t S\I{p}|$. Then, $\nabla^2 S\I{P}$, $\partial_t^2S\I{P}$ terms
also can be neglected, yielding
\begin{eqnarray}
&& 2\frac{A\I{L} A\I{P}}{c^2}+\frac{A\I{P}^2}{c^2}+2\frac{\mf{A}\I{P}}{c}\nabla S\I{L} \label{Sp}\\
&&=\frac{2}{c^2}\partial_t S\I{L}\partial_t S\I{P}+\Bigl[-2\nabla S\I{L}+\frac{2}{c}(\mf{A}\I{L}+\mf{A}\I{P})\Bigr]\nabla S\I{P}\, .\nonumber
\end{eqnarray}
The equations for the characteristics of the first order partial differential equation \eqref{Sp} are
\begin{eqnarray}
 \frac{\partial t(u)}{\partial u}&=&\frac{2}{c^2}\partial_t S\I{L}\label{eq:eik2t}\\
\frac{\partial {\bf r}(u)}{\partial u}&=&-2\nabla S\I{L}+\frac{2}{c}(\mf{A}\I{L}+\mf{A}\I{P}).\label{eq:eik3r}
\end{eqnarray}
 It follows from~Eq.~\eqref{eq:eik2t} and~Eq.~\eqref{eq:eik3r} that
\begin{equation}
 -2k\I{L}\cdot p=\mf{k}\I{L}\frac{\partial \mf{r}}{\partial u}-\omega\I{L} \frac{\partial t}{\partial u}
\end{equation}
and 
\begin{equation}
 u=-(\mf{k}\I{L}\mf{r}-\omega\I{L} t)/(2k\I{L}\cdot p)=\eta/(2k\I{L}\cdot p)\, .\label{eq:defu}
\end{equation}
Integrating~Eq.~\eqref{Sp} and employing~Eq.~\eqref{eq:defu} we derive
\begin{eqnarray}
 S\I{P} (x,x\I{0})\!\!\!&=&\!\!\!\int_{\eta_0}^\eta \!\!\!\!\D\tilde{\eta} \frac{1}{k\I{L}\cdot p}\Bigl[\nabla S\I{L}+\frac{\mf{A}\I{L}(\tilde{\eta})}{c}+\frac{\mf{A}\I{P}(\tilde{\eta},\tilde{\mf{x}}\I{L}(\mf{x},\tilde{\eta},\eta))}{2c}\Bigr]\nonumber\\
&&\;\;\;\;\times\mf{A}\I{P}(\tilde{\eta},\tilde{\mf{x}}\I{L}(\mf{x},\tilde{\eta},\eta))/c\, 
\end{eqnarray}
with
\begin{eqnarray}
\tilde{\mf{x}}\I{L}(\mf{x},\tilde{\eta},\eta)&=& \mf{x}-\int_{\tilde{\eta}}^\eta \D\eta'\frac{1}{k\I{L}\cdot p}\Bigg\{\mf{p}+\frac{\mf{A}\I{L}(\eta')}{c}+\frac{\mf{k}\I{L}}{k\I{L}\cdot p} \nonumber\\
&&\;\;\;\;\;\;\;\;\;\;\;\;\;\;\;\times\Bigl[p_x+\frac{A\I{L}(\eta')}{2c}\Bigr]\frac{A\I{L}(\eta')}{c}\Bigg\} \label{ztraj}\, .
\end{eqnarray}
The integral in Eq.~\eqref{ztraj} can be omitted in the case if the z-dependence of $A\I{P}$ along the classical trajectory of the particle in the laser field is negligible. 
For the total phase of the wave function, we therefore find
\begin{eqnarray}
 S\I{T}(x,x\I{0})&=&S\I{L}(x,x\I{0})+S\I{P}(x,x\I{0})\nonumber\\
&=&\,p\cdot (x-x_0) \label{eq:rel_pertS}\\ \nonumber
\!\!\!\!\!\!\!\!\!\!\!\!\!\!\!\!&+&\int_{\eta_0}^\eta \D\tilde{\eta}\frac{\left[\mf{p}+\frac{\mf{A}(\tilde{\eta},\tilde{\mf{x}}\I{L}(\mf{x},\tilde{\eta},\eta))}{2c}\right]\frac{\mf{A}(\tilde{\eta},\tilde{\mf{x}}\I{L}(\mf{x},\tilde{\eta},\eta))}{c}}{k\I{L}\cdot p}\, ,
\end{eqnarray}
where $\eta_0$ and $x_0$ are arbitrary constants. 
We derive the propagator as follows
\begin{equation}\label{eq:relHH_prop}
  G\I{T}(x,x')\approx-\imag\,\theta(t-t')\int 
  \frac{c \,\D^3\mathbf{q} }{2\varepsilon_{\mathbf{q}} (2\pi)^3}
  \exp\big[-\imag\,\Phi(x,x')\big]\\ \,,
\end{equation}
where $\Phi(x,x')\equiv S\I{T}(x,x\I{0})- S\I{T}(x',x\I{0})$. The latter can be represented as
\begin{eqnarray}
\Phi(x,x')=S\I{T}(x,x')+ \Delta\I{T}(x,x')\, 
\end{eqnarray}
with
\begin{eqnarray}
&& \Delta\I{T}(x,x')\nonumber\\
&=&\frac{1}{k\I{L}\cdot p}\int_{\eta_0}^\eta \D\tilde{\eta}\bigg\{\left[\mf{p}+\frac{\mf{A}(\tilde{\eta},\tilde{\mf{x}}\I{L}(\mf{x},\tilde{\eta},\eta))}{2c}\right]\frac{\mf{A}(\tilde{\eta},\tilde{\mf{x}}\I{L}(\mf{x},\tilde{\eta},\eta))}{c} \nonumber\\
&&- \left[\mf{p}+\frac{\mf{A}(\tilde{\eta},\tilde{\mf{x}}\I{L}(\mf{x},\tilde{\eta},\eta'))}{2c}\right]\frac{\mf{A}(\tilde{\eta},\tilde{\mf{x}}\I{L}(\mf{x},\tilde{\eta},\eta'))}{c} \bigg\}\, .\label{eq:Delta-eikonal}
\end{eqnarray}
The two terms in the integrand of Eq.~\eqref{eq:Delta-eikonal} deviate only in $\tilde{\mf{x}}\I{L}(\mf{x},\tilde{\eta},\eta)$  and $\tilde{\mf{x}}\I{L}(\mf{x},\tilde{\eta},\eta')$. Within the saddle-point approximation applied later, $\eta$ and $\eta'$ are the phase of recollision and ionization, respectively. Therefore, $\tilde{\mf{x}}\I{L}(\mf{x},\tilde{\eta},\eta)$  and $\tilde{\mf{x}}\I{L}(\mf{x},\tilde{\eta},\eta')$ differ only by the distance in space between ionization and recollision which is zero. Thus, we can omit $\Delta\I{T}(x,x')$.

\section{Relativistic phase-matched x-ray assisted HHG}\label{sec:relxray}

\begin{figure}[b]
\begin{center}
 \includegraphics[width=0.45\textwidth]{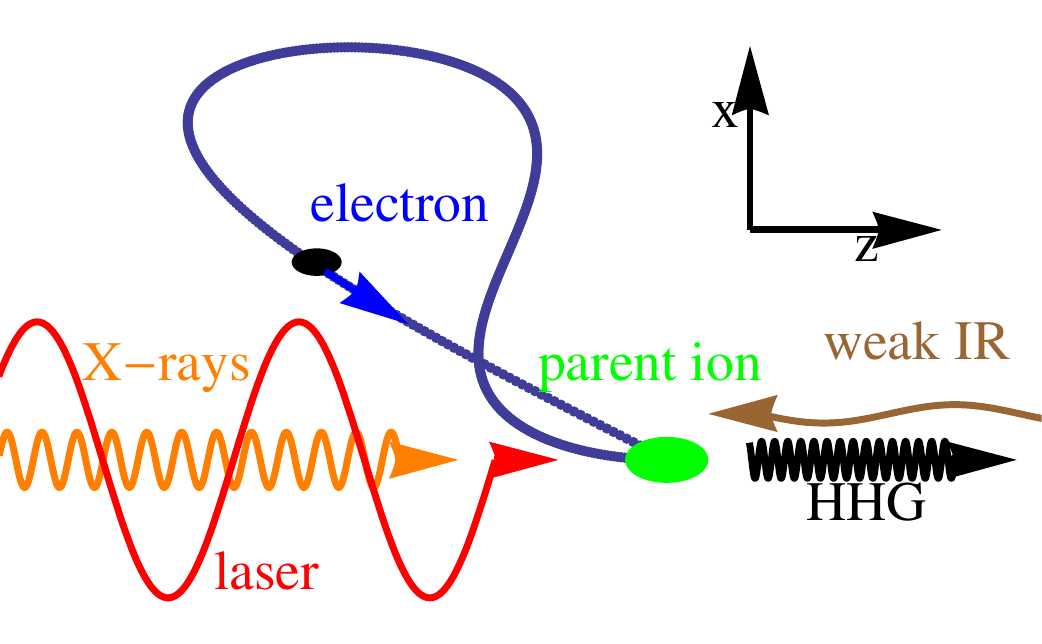}
\caption{(color online). Geometry of the HHG process for a collinear alignment of the x-ray and laser field. The copropagating x-ray field (orange) has a frequency above the ionization energy to achieve drift compensation. The weak IR field (brown) is employed to accomplish phase-matching.} \label{xray_setup2}
\end{center}
\end{figure}

In this section, the theory developed in Sec.~\ref{sec:relHHG_macr} is employed to investigate macroscopic harmonic emission in the relativistic regime for a 
HHG setup where an IR laser field of relativistic intensity is assisted by an x-ray field.
In this scheme, the x-ray frequency $\omega\I{X}$ exceeds  the binding energy $I\I{p,x}$ of the electron and thus delivers an initial momentum to the freed electron which can balance the subsequent drift motion. 
The laser alignment of the setup and an example of a classical trajectory that recollides are illustrated in  Fig.~\ref{xray_setup2}. The weak counterpropagating IR field (brown) is important for a phase-matched macroscopic response and can be ignored when discussing the process for a single atom.
The HHG medium is a macroscopic gas of multiply-charged ions.

\subsection{Current density}\label{sec:current density}
We begin with adapting the general equation for the current density Eq.~\eqref{j_expr_L} to the present setup.
Presuming that for the realization of phase-matching additional weak fields $\mf{A}\I{P}(\eta,\mf{x})$
will be required, we employ the Green function $G\I{T}(x,x^{\prime})$ of Eq.~\eqref{eq:relHH_prop} instead of the Volkov propagator
$G\I{L}(x,x^{\prime})$ in the general expression for the current density of Eq.~\eqref{j_expr_L}: 
\begin{eqnarray}
 \tilde{j\I{a}}(\mf{x\I{a}},\omega\I{H},\mf{n}')\nonumber\\
&&\!\!\!\!\!\!\!\!\!\!\!\!\!\!\!\!\!\!\!\!=\frac{2}{c} \int \D^4 x\int \D^4 x'\euler^{\imag\omega\I{H} t-\imag\mf{k}_H\mf{r}}\phi^*(\mf{x}-\mf{x}\I{a})\nonumber\\
&&\!\!\!\!\!\!\!\!\!\!\!\!\times(\hat{\mf{j}}-\tfrac{1}{2} \mf{k}\I{H})G\I{T}(x,x')V\I{I}(x')\phi(\mf{x}'-\mf{x}\I{a}) \label{eq:current-int}\, .
\end{eqnarray}
In this expression, the x-ray field enters only into the potential 
$V\I{I}(x)=2\,\mf{x}\cdot (\mf{E\I{X}}+\mf{E\I{L}})$ in the Klein-Gordon formalism.
Ignoring tunnel ionization by the IR laser field which is justified in the
considered setup, we drop the laser field in this term and approximate $V\I{I}(x) \approx 2\,\mf{x}\cdot
\mf{E\I{X}}\approx \mf{x}\cdot \mf{E}_{X0} \euler^{-\imag(\omega\I{X} t-
\mf{k}\I{X}\cdot \mf{x})}$. In the last step, the exponential function with the positive argument is dropped because
 it leads to an unphysical solution of the saddle point equations~\cite{FARIA:HO-07}.
Due to its negligible ponderomotive potential, $\mf{E\I{X}}$ can be neglected for the continuum propagation of the electron and, thus, it is neglected in the propagator.

The dependence of the current density Eq.~\eqref{eq:current-int} on the position of the atom $\mf{x}\I{a}$ is given by the bound wave functions $\phi(\mf{x}-\mf{x}\I{a},t)$. To separate out phase factors that highly oscillate with $\mf{x}\I{a}$, a coordinate transformation is applied: $\mf{\tilde{x}}=\mf{x}-\mf{x}\I{a}$. Thereafter, time integration is transformed to an integration over the laser phase: $\eta=\omega\I{L} t- \mf{k}\I{L}\cdot \mf{x}=\omega\I{L} t- \mf{k}\I{L}\cdot \mf{x}\I{a}- \mf{k}\I{L}\cdot \mf{\tilde{x}}$.
Finally, we obtain an expression for the current density that can be evaluated within the saddle-point approximation:
\begin{eqnarray}
 &&\tilde{\mf{j}}\I{a}(\mf{x}\I{a},\omega\I{H},\mf{n}') =\int^{\omega\I{L}T\I{P}}_{0}\D\eta\int^{\eta}_{-\infty}\D\eta'\int \D^3\mathbf{q}\,
  m^j(\mathbf{q},\eta,\eta',\mf{x}\I{a}) \nonumber\\
  &&\times \exp\left[-\imag( \tilde{S}\I{P}(\mathbf{q},\eta,\eta',\mf{x}\I{a})+\frac{\omega\I{X}}{\omega\I{L}} \eta'-\frac{\omega\I{H}}{\omega\I{L}}\eta)\right]\nonumber \\
&&\times \exp\left[\imag(\frac{\omega\I{H}}{\omega\I{L}}\mf{k}\I{L}-\mf{k}\I{H}+\mf{k}\I{X}-\frac{\omega\I{X}}{\omega\I{L}}\mf{k}\I{L})\mf{x}\I{a}\right]
\label{current_dens_ass}
\end{eqnarray}
where $T\I{P}$ is the laser pulse duration and 
\begin{eqnarray}
 &&m^j(\mathbf{q},\eta,\eta',\mf{x}\I{a})=-\imag\frac{c^2 (\mathbf{q}+\frac{\mathbf{A}(\eta,\mf{x}\I{a})}{c}-\frac{{\bf
k}\I{L}}{\omega\I{L}}\varepsilon_{{\bf q}} -\tfrac{1}{2} \mf{k}\I{H}  )}
{\varepsilon_\mathbf{q}\omega\I{L}^2} 
\nonumber\\&&\times  \left<0|\mathbf{q}+\frac{\mathbf{A}(\eta,\mf{x}\I{a})}{c}-\frac{{\bf k}\I{L}}{\omega\I{L}}(\varepsilon_{{\bf q}}+I\I{p,x}-c^2)+\frac{\omega\I{H}}{\omega\I{L}}\mf{k}\I{L}-\mf{k}\I{H}\right>  \nonumber \\
&&\times \left<\mathbf{q}+\frac{\mathbf{A}(\eta^{\prime },\mf{x}\I{a})}{c}-\frac{{\bf k}\I{L}}{\omega\I{L}}(\varepsilon_{{\bf q}}+I\I{p,x}-c^2)  
\right.\nonumber\\ && \;\;\;\;\;\;\;\;\;\;\;\;\;\;\;\;\;\;\;\;+ \left.
\frac{\omega\I{X}}{\omega\I{L}}\mf{k}\I{L}-\mf{k}\I{X}\left|\mf{x}\cdot \mf{E}_{X0}\right|0\right>\, ; \label{mh5}
\end{eqnarray}
Further,
\begin{equation}\label{eq:tildeSpert}
\tilde{S}_{\mathrm P}(\mathbf{p},\eta,\eta',\mf{x}\I{a})=
\int^{\eta}_{\eta'}
d\tilde{\eta}\left(\tilde{\varepsilon}_{\mathbf{q}}^{\mathrm P}
  (\tilde{\eta},\tilde{\mf{x}}(\mf{x}\I{a},\tilde{\eta},\eta))-c^2+I\I{p,x}\right)/\omega\I{L}  
\end{equation}
with the relativistic energy of the electron in the position dependent laser field given by
\begin{eqnarray}
  \tilde{\varepsilon}_{\mathbf{q}}^{\mathrm P}(\eta,\mf{x})=\varepsilon_{\mathbf{q}}
  +\frac{\omega\I{L}}{k\I{L}\cdot q}  
\left(\mathbf{q}+\frac{\mathbf{A}(\eta,\mf{x})}{2c}\right)\cdot\frac{\mathbf{A}
(\eta,\mf{x})}{c}\, .
\end{eqnarray}
The wave vector of the laser is $\mf{k}\I{L}=n\I{ref}\frac{\omega\I{L}}{c} \hat{\mf{e}}\I{z}$ and 
\begin{equation}
 n\I{ref}=\sqrt{1-\frac{\omega\I{p}^2}{\omega\I{L}^2}}\label{eq:plasma_n}
\end{equation}
is the refractive  index of the plasma, with the plasma frequency $\omega\I{p}=\sqrt{4\pi Z\rho}$ and the ion charge number $Z$. 
This way, we take into account the change of the phase velocity caused by the free electrons but ignore pulse deformation.
Further, we restrict ourselves to a weakly focused laser field which can be treated as a plane wave and ignore the transversal variation of the vector potential  with respect to the propagation direction. 

The propagation direction and the frequency of the assisting x-ray field
$\mf{E\I{X}}$ have to be chosen in such a way to facilitate the phase-matching
and to counteract the relativistic drift effectively. Let us first consider the
feasibility of phase-matching by analyzing the harmonic emission phase
of different ions in the medium via Eq.~\eqref{current_dens_ass}.  Generally,
the vector potential consists of two terms as in section~\ref{sec:eikonal} 
\begin{equation}\label{eq:totalA} 
\mf{A}(\eta,\mf{x})=\mf{A}\I{L}(\eta)+\mf{A}\I{P}(\eta,\mf{x}).
\end{equation}

To understand the most favorable condition for phase-matching, let us assume for a moment that there are no additional fields for facilitating quasi-phase matching and ignore deformation of the laser pulse. Then, we can omit $\mf{A}\I{P}(\eta,\mf{x})$ in~Eq.~\eqref{eq:totalA} 
[a modified setup where the additional $\mf{A}\I{P}(\eta,\mf{x})$ field is taken into account will be considered in Sec. \ref{sec:macro}]. 
In this case, we have no $\mf{x}_A$ dependence of
$m^j(\mathbf{q},\eta',\eta'',\mf{x}_A)$ and $\tilde{S}\I{P}(\mathbf{q},\eta',\eta'',\mf{x}_A)$.
The only contribution to the variation of the harmonic emission phase along the medium comes from the exponential function $\exp(\imag \Delta \mf{k}\, \textbf{r}_a)$  in~Eq.~\eqref{current_dens_ass} determining the coherence length $l_{coh}=\pi/|\Delta \mf{k}|$, with the phase-mismatching wave-vector 
\begin{equation}
\Delta\mf{ k}=\frac{\omega\I{H}}{\omega\I{L}}\mf{k}\I{L}-\mf{k}\I{H}+\mf{k}\I{X}-\frac{\omega\I{X}}{\omega\I{L}}\mf{k}\I{L}\, ,\label{eq:xrayass_mismatch}
\end{equation}
caused by the refractive index difference between the fields. The phase-mismatch Eq.~\eqref{eq:xrayass_mismatch} can be split up in two parts. One is due to the phase-mismatch between the harmonics and the laser and the other one due to the phase-mismatch between the ionizing x rays and the laser. Interestingly, due to the different signs both can partially cancel out. This fact also holds true for this setup in the non-relativistic regime. Note that $\frac{\omega\I{H}}{\omega\I{L}}\mf{k}\I{L}$ and $\mf{k}\I{H}$ are much larger than the other two terms left in~Eq.~\eqref{eq:xrayass_mismatch} for the case $\omega\I{H}\gg\omega\I{X}$. The phase mismatch is smallest in the case of a collinear alignment of the x-ray and laser fields $\mf{k}\I{H}\parallel \mf{k}\I{L}$ (see Fig.~\ref{xray_setup2}) with propagation in the z direction. 
Accordingly, only the x component of the spectral current density of Eq.~\eqref{current_dens_ass} will contribute to the harmonic emission in the z direction. 

The drift compensation can be achieved with an appropriate choice of the x-ray frequency which is considered in Sec.~\ref{sec:relHHGass_saddlepoints}.

\subsection{Single-atom response}\label{sec:single-atom}

\subsubsection{Singe-atom HHG rate}

The single-atom photon emission rate per solid angle of  the $n^{{\rm
th}}$ harmonic can be calculated from
\begin{eqnarray}
\frac{dw_n}{d\Omega}&=& \frac{1}{T\I{P}} \int_{n \omega\I{L}-\omega\I{L}}^{n \omega\I{L}+\omega\I{L}}\D \omega\I{H} \frac{\D N}{\D \omega\I{H} \D\Omega} \\ \nonumber
&=& \frac{c R^2}{(2\pi)^2\omega\I{H} T\I{P}} \int_{n \omega\I{L}-\omega\I{L}}^{n \omega\I{L}+\omega\I{L}}\D \omega\I{H}  \vert\tilde{\mf{E}}\I{H,0}(\mf{n}',\omega\I{H}) \vert^2\, ,
\label{mhhg3a}
\end{eqnarray}
using Eq.~\eqref{phase_match_integralEx} for
$\tilde{\mf{E}}\I{H,0}(\mf{n}',\omega\I{H})$ and the density
$\rho(\textbf{x}\I{a})=\delta(\textbf{x}\I{a})$ which yields
\begin{equation}
\frac{dw_n}{d\Omega}=  \frac{\omega\I{L}^2 \omega\I{H}}{(2 \pi c)^3 } \vert\tilde{\mf{j}}\I{a}(0,n \omega\I{L},\mf{n}') \vert^2\, ,
\label{mhhg4a}
\end{equation}
where in the expression for $\tilde{\mf{j}}\I{a}(0,n \omega\I{L},\mf{n}')$ given by Eq.~(\ref{current_dens_ass}), $T\I{P}$ is replaced by $\frac{2\pi}{\omega\I{L}}$ to confine the emission to one laser period [the rate of Eq.~(\ref{mhhg4a}) is identical to ones in Refs.~\cite{HATSAGORTSYAN:UFO-08,KLAIBER:CH-08}].

\begin{figure}[h]
\begin{center}
 \includegraphics[width=0.4\textwidth]{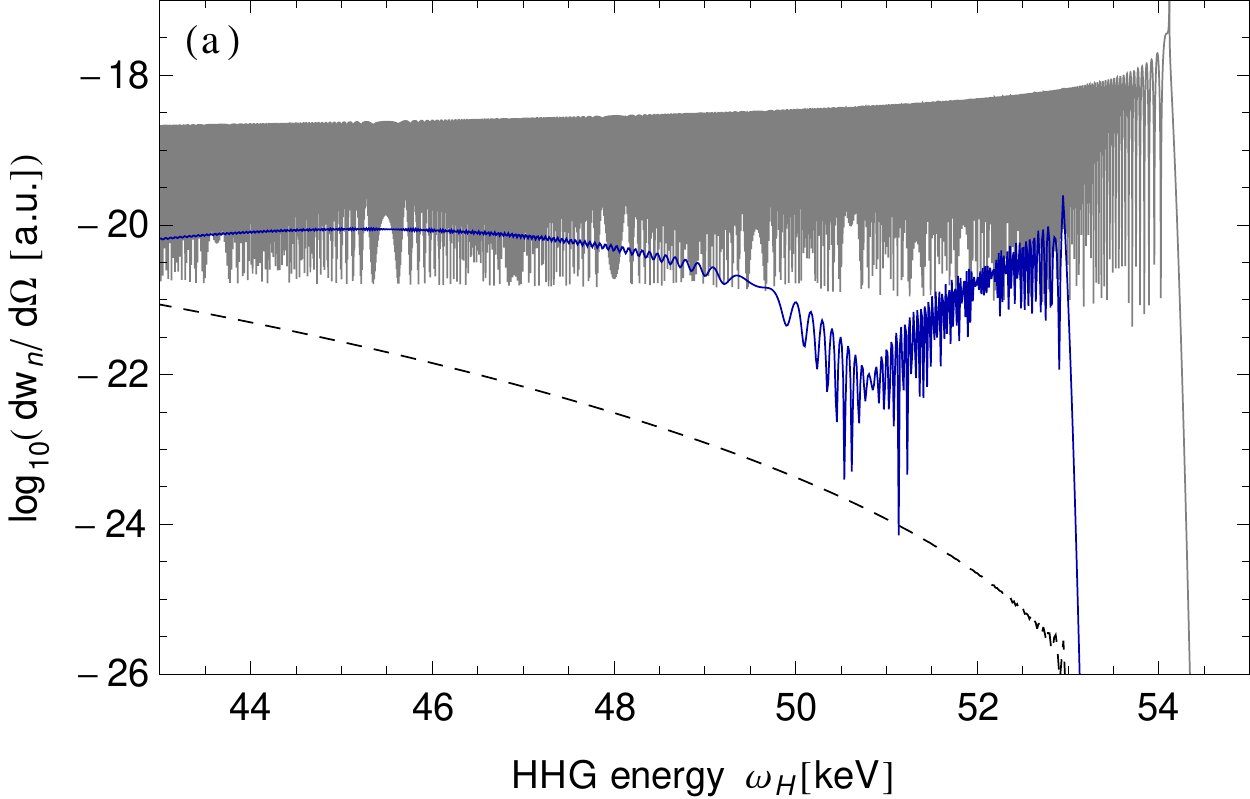}
\vskip0.2cm
 \includegraphics[width=0.4\textwidth]{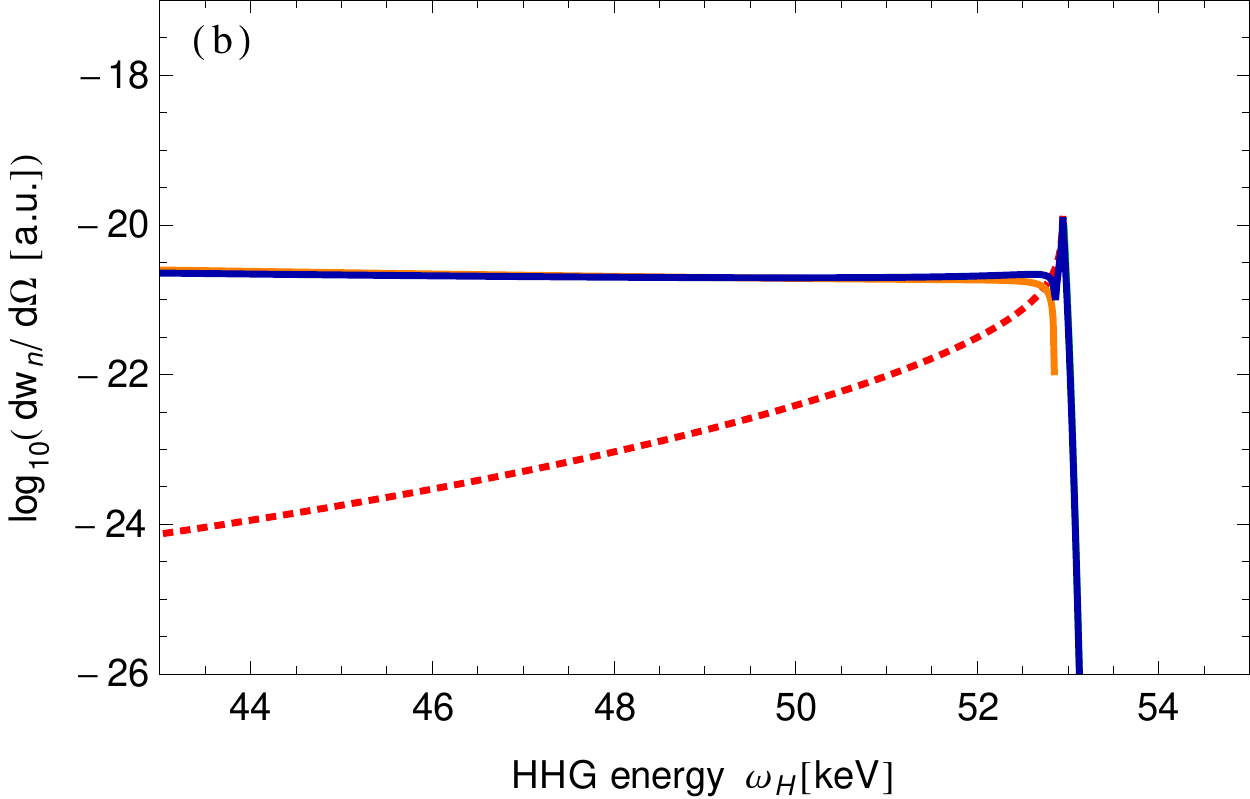}
\caption{(color online). (a) Single-atom emission probability for $E_0=2.5\au$, $I\I{p,x}=8\au$, $\omega\I{X}=14\au$ and $E\I{X}=0.65\au$ (blue, dark gray) and a conventional laser field ($E\I{X}=0$) with $E_0=2.5\au$   and $I\I{p,t}=4.8\au$ (dashed black) and the same configuration in the DA (gray). For the second configuration, $I\I{p,t}$ is chosen such that the average tunnel-ionization rate is the same as the single-photon ionization rate in the case before. (b) Separate HHG yields of the three contributing quasi-classical trajectories for the discussed setup [blue line in (a)]. The dotted red contribution (long trajectory) is suppressed because the drift for the trajectory is not completely compensated (as discussed in Sec.~\ref{sec:relHHGass_saddlepoints}). The short trajectory contributions (solid lines in dark blue, dark gray and orange, gray) are nearly identical.
} \label{fig:single_atom_xray}
\end{center}
\end{figure}

A typical HHG spectrum for the considered setup calculated within the
saddle-point approximation is displayed in Fig.~\ref{fig:single_atom_xray}~(a)
in blue for the set of parameters denoted in the caption of the figure. In this
paper we employ a zero-range potential~\cite{BECKER:ZP-94} to model the
binding potential.  
Analytical expressions of the matrix elements which appear 
in Eq.~\eqref{mh5} can be found in, e.g.,~\cite{MILOSEVIC:RL-02,KLAIBER:LO-07}.
The ionization potential $I\I{p,x}=8\au$ is chosen large enough such that tunnel ionization by the strong optical laser field does not lead to depletion of the bound wave function.  We compare the spectrum obtained from the x-ray assisted setup (blue line) with the spectrum of a
conventional HHG setup where no x-ray field is present, calculated either fully
relativistically~\cite{KLAIBER:LO-07} (dashed black) or within the
dipole approximation (DA)~\cite{MILOSEVIC:RL-02} (gray).
The ionization matrix elements for the conventional HHG setups was multiplied by the factor
$\frac{2\sqrt{2}\kappa\I{t}^3}{\vert E\I{L}(t') \vert}$~\cite{MILOSEVIC:RT-02} to account for the underestimation of the tunneling rate when employing the zero-range potential. Here, $t'$ is the ionization time and $\kappa\I{t}=\sqrt{2I\I{p,t}}$, with $I\I{p,t}$ being the ionization potential in the case of the conventional setup.
In order to have a fair comparison of  the x-ray assisted setup with the conventional one, we have to choose the ionization potential of the conventional setup ($I\I{p,t}$) different from that of the x-ray assisted one ($I\I{p,x}$) such that the ionization rates of both setups were the same. 
For this purpose, the ionization rate for the conventional setup is calculated from the Perelomov, Popov, Terent'ev (PPT) tunneling rate~\cite{PERELOMOV:IA-67,AMMOSOV:TR-86}, while the single-photon ionization rate from the zero-range potential is derived using the differential photoionization cross section in nonrelativisitic, dipole approximation and approximating the ionized electron wave function by a plane wave (in analogy to Ref.~\cite{bethe1968intermediate}):
\begin{eqnarray}
 \frac{\D \sigma\I{x,1ph}}{\D \Omega}=\frac{p}{2\pi c \omega\I{X}}\vert\langle
\mf{p}\vert z\vert 0\rangle \vert^2=\frac{2^4 p^3\omega\I{X} \kappa\I{z}}{c}\frac{\cos^2 \theta}{(\kappa\I{x}^2+p^2)^4}\label{eq:diff_1-photon}
\end{eqnarray}
where  $\theta$ is the angle between the electron momentum $\mf{p}$ and the polarization direction $\hat{\mf{z}}$ of the x-ray field. An integration over all emission angles yields the total cross section
\begin{equation}\label{eq:io-rate-zr}
 \sigma_{\delta,1ph}=\frac{4 \kappa\I{x}\pi}{3c}\frac{(2\omega\I{X}-\kappa\I{x}^2)^{3/2}}{\omega\I{X}^3}\, ,
\end{equation}
where we used $p=\sqrt{2\omega\I{x}-\kappa\I{x}^2}$.

The main message of Fig.~\ref{fig:single_atom_xray}~(a) is that the relativistic
drift can be fully compensated in the x-ray assisted HHG setup
(in the case of the chosen x-ray frequency $\omega\I{X}-I\I{p,x}= 6$ a.u.,
the gray and blue curve are of comparable order, small suppression arises from the different spreading behavior). The yield of the considered setup is much higher than that for the conventional setup (dashed black), the latter being suppressed by the drift.

In the next section, we explain how the single-x-ray-photon ionization provides the necessary initial momentum for the electron (opposite to the IR laser propagation direction) to counteract the relativistic drift in the case when the x-rays propagates along the strong IR laser field. We discuss also the optimization of the applied x-ray frequency for the HHG process.

\subsubsection{Drift compensation and influence of x-ray frequency}\label{sec:relHHGass_saddlepoints}
The integration in~Eq.~\eqref{current_dens_ass} is carried out via the
saddle-point integration method~\cite{Arfken:MM-05,Lewenstein:HH-94}. This
means that instead of the integration we only need to sum the integrand over a
small number of saddle points for each energy $\omega\I{H}$. A saddle point
$(\eta,\eta',\mf{q})$ determines the ionization and recollision times and
the canonical momentum for the  electron classical trajectory leading to the
harmonic energy under consideration. In general, they are complex expressing
non-classical dynamics during tunneling ionization.
For the parameters chosen above, 3 quantum paths (saddle points) contribute to the spectrum for each energy  being equivalent to three classical trajectories that recollide with that energy. The separate contributions of each quantum path to the spectrum are shown in Fig.~\ref{fig:single_atom_xray}~(b). The two paths marked in blue and orange (solid lines) have nearly the same yield whereas the dotted red line is suppressed by several orders of magnitude. In the following we explain the reason for the difference and discuss the influence of the x-ray frequency on the dynamics.

\begin{figure}[h!] \begin{center} 
\includegraphics[width=0.45\textwidth]{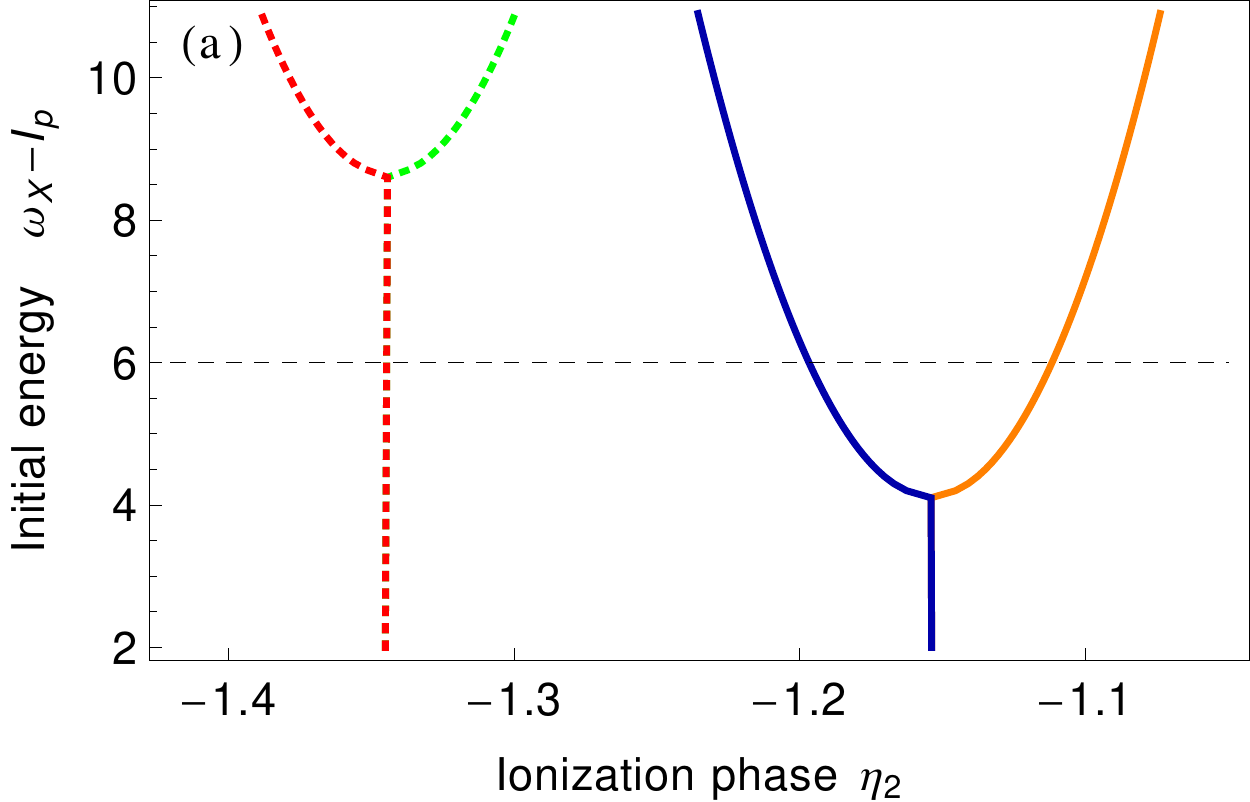}
\vskip0.2cm \includegraphics[width=0.45\textwidth]{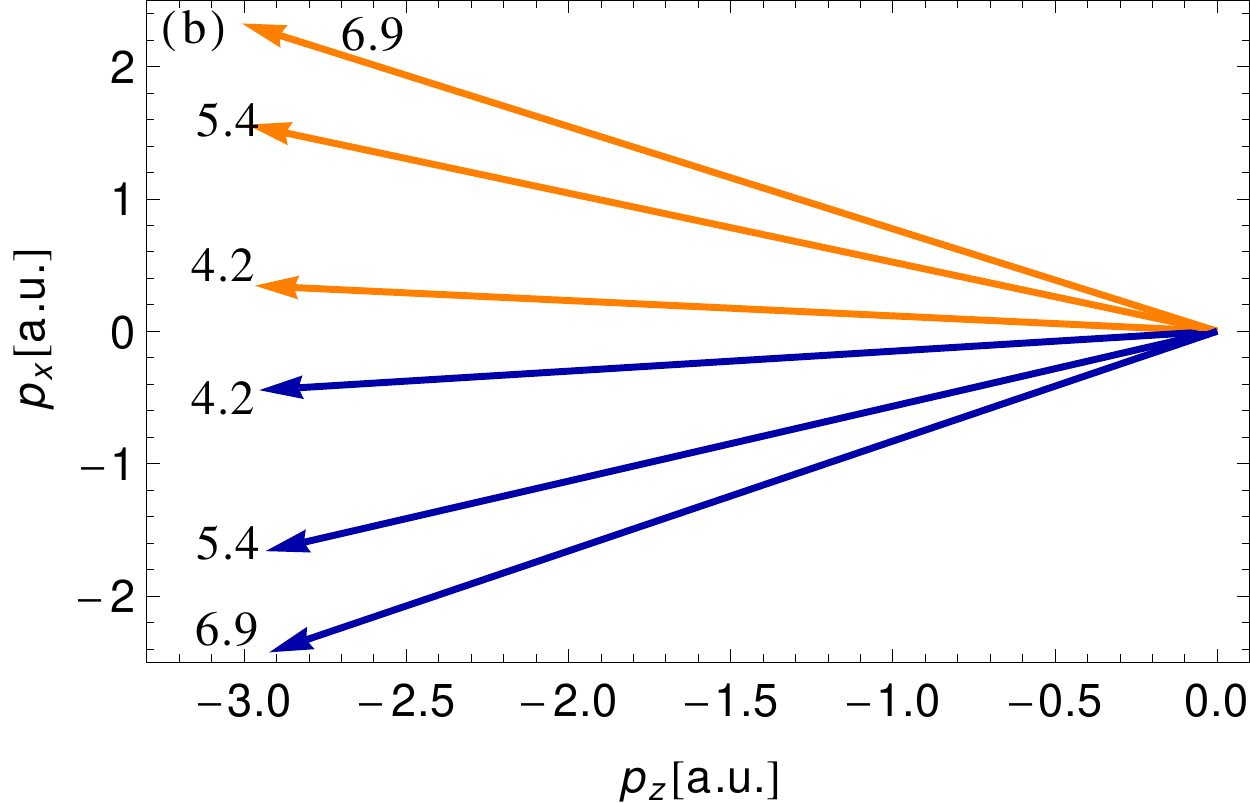}
\vskip0.2cm\includegraphics[width=0.45\textwidth]{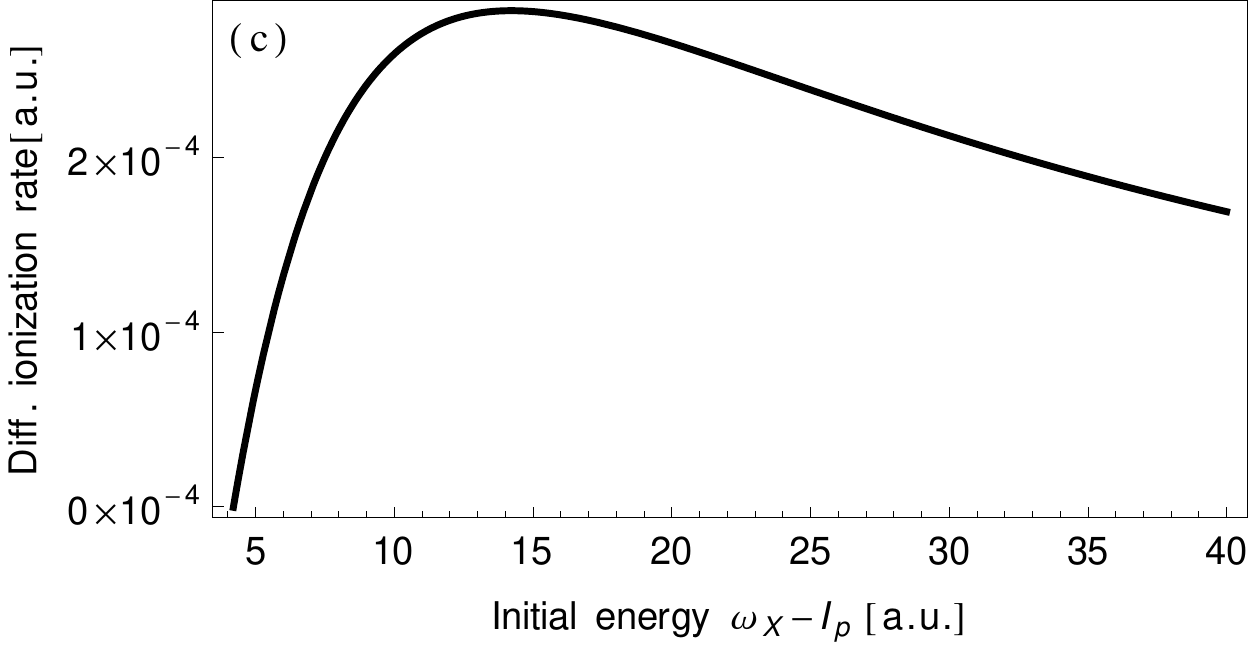}
\caption{(color online). (a) We show the ionization phase saddle points of the $50\U{keV}$ trajectory for different x-ray frequencies which is the same as ionization time in units of radian. The dashed line indicates the value chosen in Fig.~\ref{fig:single_atom_xray}. The dotted and solid lines belong to the long and short trajectories, respectivly.  (b) displays the initial momentum direction for different initial energies $\omega\I{X}-I\I{p,x}$ [as indicated in a.u. next to the respective arrow] needed for the emission of a $50\U{keV}$ photon. The uper and lower branch  correspond to the short up- and downhill  trajectories, respectivly, where the color indication coincides with the one in (a). Note that the ionization phases $\eta_2$ are different for the up- and downhill trajectories. $\hat{\mf{x}}$ and $\hat{\mf{z}}$ are the propagation and polarization directions of the laser, respectively. (c) Differential ionization rate depending on the initial energy $\omega\I{X}-I\I{p,x}$ from Eq.~\eqref{eq:diff_1-photon}. The considered direction is  in the initial momentum direction determined by the saddle point equations. From (b) we see that the z-component is approximately $p\I{z,c}=2.9\au$ and the x-component depends on $\omega\I{X}-I\I{p,x}$.} \label{fig:xray_ass_wx-dep}
\end{center}
\end{figure}

In order to understand the number of contributing trajectories in Fig.~\ref{fig:single_atom_xray}~(b), we calculate the saddle-point solutions for different x-ray frequencies $\omega\I{X}$ for the harmonic emission at $50\U{keV}$ and show the ionization phase saddle point in Fig.~\ref{fig:xray_ass_wx-dep}~(a). For small  initial energies $\omega\I{X}-I\I{p,x}$, two saddle points contribute to harmonic emission as in the usual case of HHG in a laser field only. Both saddle points, the long ($\mathrm{Re}\,\eta_2\approx -1.345$) and short ($\mathrm{Re}\,\eta_2\approx -1.115$) trajectory, are complex [their real part  is shown in the graph] and their HHG amplitude is very tiny due to the missing drift compensation which is indicated by the complex value. When increasing $\omega\I{X}$, first the short trajectory  and then the long trajectory  split up into two parts. These branches are called uphill and downhill trajectories, respectively, because their initial momentum component along the laser polarization is either positive or negative~\cite{FARIA:HO-07}. After the splitting at about $\omega\I{X}-I\I{p,x}\approx4$ a.u. and $\omega\I{X}-I\I{p,x}\approx 8.5$ a.u., the respective ionization phase is purely real which indicates that the initial momentum is sufficient to compensate the subsequent relativistic drift. The short trajectory reaches drift compensation earlier because it spends less time in the continuum and, therefore, undergoes a smaller drift that requires compensation.

The dashed line in Fig.~\ref{fig:xray_ass_wx-dep}~(a) denotes the x-ray frequency $\omega\I{X}$ that was chosen in Fig.~\ref{fig:single_atom_xray}~(b). The short trajectory has two contributions (blue and orange) whereas the long trajectory (dotted) has only one contribution (red). The contribution of the long trajectory is suppressed by about 3 orders of magnitude compared to the short contributions which is visible from Fig.~\ref{fig:single_atom_xray}~(b). This is  because the long trajectory spends more time in the continuum and experiences a larger relativistic drift  which cannot be fully compensated for. In this case, the ionization saddle point is complex leading to a damping in the exponential function in the respective amplitude~Eq.~\eqref{current_dens_ass}. By increasing $\omega\I{X}$ above $\omega\I{X}-I\I{p}\approx8.5$ a.u., the drift compensation could also be achieved for the long trajectory and the dashed red contribution in the spectrum could be enhanced leading to a larger single-atom yield. However, only one of the trajectories can be phase-matched in many cases and the enhancement of the other trajectories would not be useful.

In Fig.~\ref{fig:xray_ass_wx-dep}~(b), the initial momentum vectors of the
ionized electron $\mf p(\eta^{\prime},\mf{q})=\mathbf{q}+\frac{\mathbf{A}(\eta^{\prime})}{c
}-\frac{{\bf k}}{\omega\I{L}}(\varepsilon_{{\bf q}}+I\I{p,x}-c^2)$, which
correspond to solutions of the saddle-point equations, are displayed for
different $\omega\I{X}$. When $\omega\I{X}-I\I{p,x}\approx4.2\au$, the momentum
required for drift compensation of the short trajectories is just reached [see
Fig.~\ref{fig:xray_ass_wx-dep}~(a)]. In this case, the initial momentum is
directed mainly along the z-direction [arrows marked with $4.2$ in
Fig.~\ref{fig:xray_ass_wx-dep}~(b)].  When $\omega\I{X}$ is increased, only the
$p\I{x}$ component changes; the $p\I{z}$ component  approximately remains
constant because it is determined by the drift compensation condition. The
electrons with an appropriate initial momentum vector can be provided by the
x-ray single-photon photoionization because the latter happens with a large
angular distribution with a maximum around x-ray polarization direction as can
be seen from the differential ionization cross section
of~Eq.~\eqref{eq:diff_1-photon}. Because the HHG amplitude for each trajectory
contains the differential ionization cross section~Eq.~\eqref{eq:diff_1-photon},
the efficiency in each case depends on the scalar product between required
ionization direction $\mf p$ and x-ray field polarization direction. This
results in some freedom in choosing the direction of $\mf{E}\I{X}$. Only if $\mf
p$ and $\mf{E}\I{X}$ were close to perpendicular [$\theta\approx \pi/2$ in
Eq.~\eqref{eq:diff_1-photon}], the differential ionization probability would be
close to zero. For realization of phase-matching, 
as it is shown above in Sec. \ref{sec:current density},
the collinear propagation of the laser and x-ray field is advantageous. For this case of a collinear alignment, we show the differential ionization probability $\D \sigma\I{x,1ph}/\D \Omega$ from Eq.~\eqref{eq:diff_1-photon} for different x-ray frequencies $\omega\I{X}$ in Fig.~\ref{fig:xray_ass_wx-dep}~(c). The emission angle of interest is estimated by the initial momentum $p=\sqrt{2 (\omega\I{X}-I\I{p,x})}$ and its z-component $p\I{z,c}=2.9\au$ taken from Fig.~\ref{fig:xray_ass_wx-dep}~(b) via $\sin \theta=p\I{z,c}/p$.  For initial energies just above $4$ [e.g., 4.2 corresponding to the nearly horizontal vectors in Fig.~\ref{fig:xray_ass_wx-dep}~(b)], $\D \sigma\I{x,1ph}/\D \Omega$ is vanishing because the momentum direction and the direction of   $\mf{E}\I{X}$ are perpendicular. The angle of emission $\theta$ will increase with rising $\omega_X$, increasing the ionization probability. On the other side, large values for $\omega\I{X}$ decrease the overall ionization probability due to the denominator of Eq. (\ref{eq:diff_1-photon}). These two competing tendencies creates the maximum in the ionization probability in Fig.~\ref{fig:xray_ass_wx-dep}~(c). We see that the  chosen value $\omega\I{X}-I\I{p,x}=6\au$ ($\omega\I{X}=14$) is close to the optimal conditions.

The former trajectory-based discussion can also be seen from a wave packet perspective. The single-photon ionization mechanism with a large initial kinetic energy obeys a dipole angular distribution of the ejected wave packet, i.e. the wave packet has an increased spreading velocity compared to tunnel ionization. The increased spatial dimension of the recolliding wave packet is exploited to overcome the  drift.

\subsection{Macroscopic HHG emission}\label{sec:macro}

After discussing the single-atom yield of the x-ray assisted setup, we continue to elaborate on the macroscopic aspect of the emission from a gas target.

\begin{figure}[h]
\begin{center}
 \includegraphics[width=0.4\textwidth]{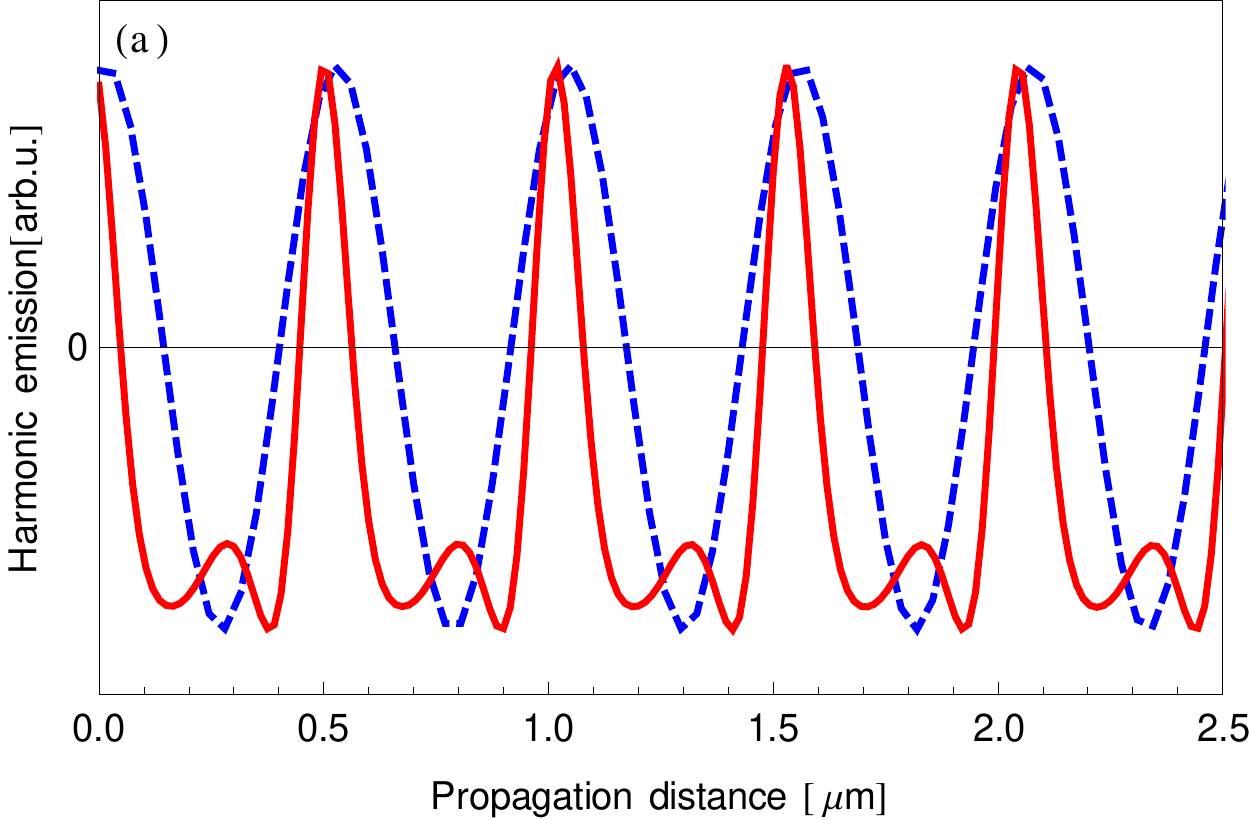}
 \includegraphics[width=0.4\textwidth]{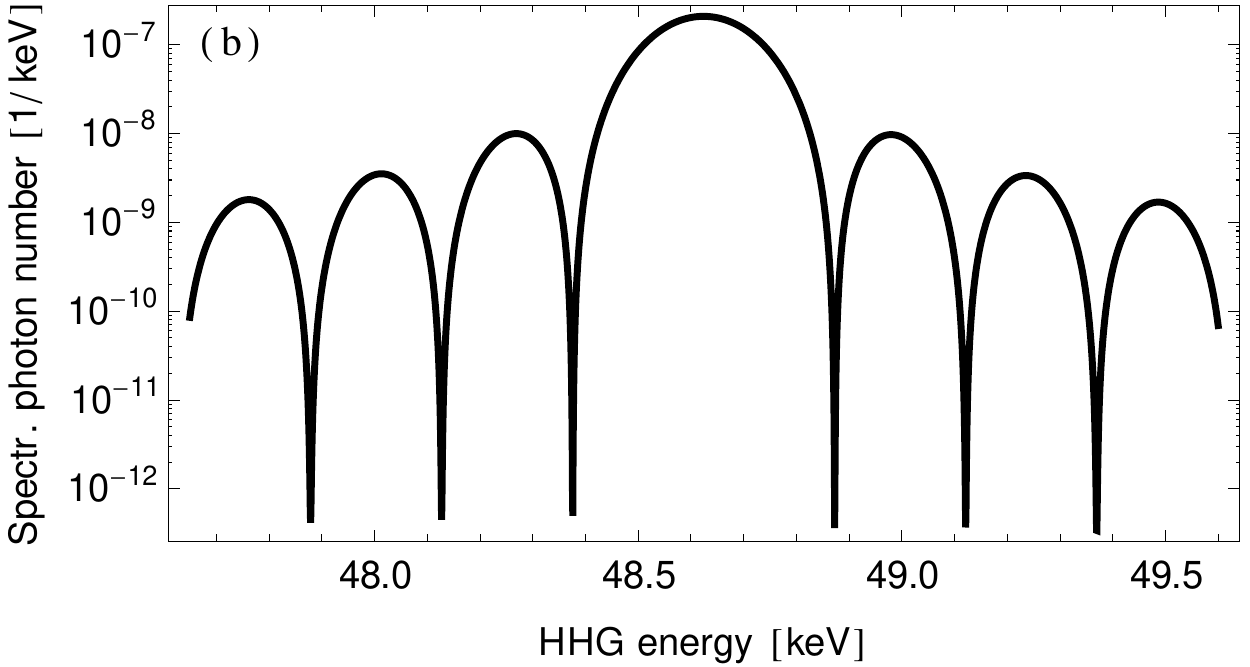}
\caption{(color online). (a) Real part of spectral component of the locally emitted HHG field at $48.6\U{keV}$ at different positions along the propagation direction. The blue dashed line is for HHG without the quasi-phase-matching scheme.  The red line is for the case of adding the weak counterpropagating field to achieve QPM. (b) The macroscopically emitted spectral photon number via Eq.~\eqref{eq:spectr-phot-num} is displayed for the QPM scenario.} \label{fig:hhgass-phase-mod-spectr}
\end{center}
\end{figure}
We inspect the emission from a  Be$^{3+}$ gas of homogeneous density $\rho=5\times10^{16}/\text{cm}^3$ with the same parameters as in Fig.~\ref{fig:single_atom_xray}. The plasma refractive index at the laser frequency  is $n\I{L}=5\times10^{-5}$. 
The phase mismatching wave-vector of Eq.~\eqref{eq:xrayass_mismatch} at the harmonic emission energy of $50\U{keV}$  is then $\Delta k=6\times10^{-4}\au$ and the coherence length  $l\I{coh}=\pi/\Delta k=0.25\,\mu \text{m}$.
In order to increase the coherence length, a quasi-phase matching (QPM) scheme can be employed~\cite{PEATROSS:SZ-97,COHEN:GA-99,SERRAT:SE-10}. 
We propose to use a weak counterpropagating IR field with the parameters $E_2=5\times10^{-5}\au$ and $\omega_2=0.0418\au$ to achieve quasi-phase matching (QPM). The additional field is denoted by a brown line in Fig.~\ref{xray_setup2}.  It is included into our mathematical formalism by $\mf{A}\I{P}(\eta,\mf{x})$ in~Eq.~\eqref{eq:totalA}. In that way a dependence on $\mf{x}\I{a}$ is introduced into $\tilde{S}\I{P}(\mathbf{q},\eta,\eta',\mf{x}\I{a})$ and the saddle points for the integration in~Eq.~\eqref{current_dens_ass} depend on the position  within the medium.  Thereby, $\tilde{\mf{x}}(\mf{x}\I{a},\tilde{\eta},\eta)$ in~Eq.~\eqref{eq:tildeSpert} contains the variation of the weak field seen by the electron along the z-direction. For the chosen set of parameters, the approximation of $\tilde{\mf{x}}(\mf{x}\I{a},\tilde{\eta},\eta)$
by $\mf{x}\I{a}$ does not lead to a significant change of the final results and, thus, can be done to save computation time. The impact of the additional field can be observed in Fig.~\ref{fig:hhgass-phase-mod-spectr}~(a). The real part of the spectral current density Eq.~\eqref{current_dens_ass} at the respective position is shown. The emitted total field is given by a spatial integral over all contributions of the current density [see Eq.~\eqref{phase_match_integral}]. 
Without QPM (blue dashed line), the single-atom contributions oscillate on the scale of
the coherence length estimated previously. An integration over all contributions
results in extensive cancellation. However, when applying the additional field,
the symmetry between the positive and negative contributions is broken (see the
red line)
and both parts only partially cancel thus achieving quasi-phase-matching and a nonzero value of the integral. The parameters of the additional field  were chosen to optimize the photon energy at $48.6\U{keV}$. A medium length of 100~$\mu$m was chosen whereas the diameter is 500~$\mu$m. The assumed laser and x-ray pulse duration is 10 cycles. The length is limited due to the assumed bandwidth of the weak QPM field $\Delta \omega_2 \sim 0.1\%$.
The spectrum is shown in Fig.~\ref{fig:hhgass-phase-mod-spectr}~(b) and an integral over the spectrum yields the final result of $5\times10^{-7}$ emitted photons per shot. The number is of similar order of magnitude as in the other relativistic HHG setup based on the counterpropagating attosecond pulse trains for driving harmonics~\cite{KOHLER:HX-11}.

\subsection{Efficiency analysis}\label{sec:relHHGefficiency}

We continue with a discussion about the small HHG yield in the relativistic regime and identify several reasons for it that are either general to the relativistic regime or specific to this setup. First, we specify an estimate expression for the emitted photon number:
\begin{equation}
N=\frac{\D w_n}{\D \Omega} \times \Delta n \times \Delta\Omega \times \Delta t  \times V^2 \rho^2 .\label{estimation}
\end{equation}
It allows in a simple way to estimate the HHG yield by an order of magnitude
and  to single out the different issues influencing the HHG yield.
In~Eq.~\eqref{estimation}, $\D w_n/\D \Omega$ is the single-atom emission rate, $\Delta n$ is the number of harmonics within the phase-matched frequency bandwidth, $\Delta t$ the interaction time that is approximately the delay between both pulses, $\Delta \Omega$ the solid angle of emitted harmonics,
$V$ the volume of coherently emitting atoms [perfect phase-matching is assumed in this volume], $\rho$ the atomic density.

First, we demonstrate the usefulness of the expression by estimating the photon number for the proposed setup and show that the result of the former exact calculation can be reproduced. We estimate the terms in Eq.~\eqref{estimation} as follows: the single atom emission rate is $\D w_n/\D \Omega\approx 10^{-21}$ [see Fig.~\ref{fig:single_atom_xray}~(b)]; the phase-matched frequency bandwidth can be deduced  from Fig.~\ref{fig:hhgass-phase-mod-spectr}~(b): $\Delta \omega\I{H}\approx 0.2\U{keV}$ which gives $\Delta n\sim 10^2$; the solid angle of phase-matched emission in the far field is determined by the interference pattern of a circular aperture $\Delta \Omega\sim\pi(2\pi c/D\omega\I{H})^2\approx10^{-14}$, where $\omega\I{H}=50\U{keV}$ and medium radius $r\I{a}=5\times10^6$a.u. are assumed; the interaction time $\Delta t\approx10^3\au$ is the laser pulse duration; the volume is cylindrical $V=\pi r\I{a}^2 \Delta z\I{a}=10^{20}$;  the plasma density restricted by dispersion is $\rho=5\times10^{16}/$cm$^3$.
Taking all pieces together, the emitted photon number under phase-matched conditions is 
\begin{eqnarray}\label{eq:estcprop} 
N\I{rel}^{50\U{keV}}&=&\frac{\D w_n}{\D \Omega}{\Big|}\I{x,s} \times \Delta n \times \Delta\Omega \times \Delta t  \times V^2 \rho^2 \\&=& 10^{-21}\times10^{2}\times10^{-14}\times10^{3}\times10^{40}\times10^{-16}\nonumber \\&=&10^{-6}\nonumber 
\end{eqnarray}
in agreement with the previous accurate calculation. The subindex x stands for the x-ray assisted setup whereas s denotes that the short trajectory contribution is taken into account only.

To explain the low yield in the relativistic regime we perform the same kind of estimation for a state-of-the-art HHG experiment~\cite{Goulielmakis:SC-08} in the non-relativistic regime where $80\U{as}$ pulses were generated from harmonics below $100\eV$. We can estimate the emitted photon number in this case:
\begin{eqnarray}\label{eq:estmpy}
 N\I{non-rel}^{50\U{eV}}&=&\frac{\D w_n}{\D \Omega}{\Big|}\I{t,s} \times \Delta n \times \Delta\Omega \times \Delta t  \times V^2 \rho^2\\
&=& 10^{-16}\times10^{1}\times10^{-7}\times10^{2}\times10^{41}\times10^{-12}\nonumber \\
&=&10^{9}\nonumber 
\end{eqnarray}
The single-atom contribution $\D w_n/\D \Omega|\I{t,s}$ was calculated from~\cite{MILOSEVIC:RL-02} where we additionally inserted a correction factor accounting for the underestimation of the tunneling rate when using the zero-range potential as described in~\cite{KOHLER:HX-11}.
By comparing~Eq.~\eqref{eq:estcprop} and~Eq.~\eqref{eq:estmpy}, one observes a dramatic suppression of 15 orders of magnitude when rising the HHG energy by about 3 orders of magnitude. It arises mainly due to the single-atom yield $\D w_n/\D \Omega$, the phase-matched emission angle $\Delta\Omega$ and the gas density. 
The single-atom contribution will be investigated separately below.
The estimated solid angle emission angle decreases quadratically with the harmonic energy. This is because a smaller harmonic wavelength leads to a smaller angle of the first interference minimum. The gas density depends on the phase-matching conditions which are much more difficult to fulfill in the relativistic regime and thus the gas density is lower in this case.

In the following, we inspect  the ratio between the single-atom yields $\D
w_n/\D \Omega$ of Eq.~\eqref{eq:estcprop} and Eq.~\eqref{eq:estmpy} closer.
In each case, we concentrate on a single (short) trajectory at $50\eV$ and
$50\U{keV}$, respectively. 
Then each single-atom rate can be estimated~\cite{HATSAGORTSYAN:UFO-08,KLAIBER:CH-08}, 
\begin{eqnarray}\label{eq:1atom_rel}
 \frac{\D w_n}{\D \Omega}{\Big|}\I{x,s}&=&\frac{1}{(2\pi c)^3\omega\I{L}^2}\omega\I{H}^2 \Big\vert\sqrt{\frac{(-2\pi i)^5}{D(\mf{q},\eta,\eta')}}\Big\vert^2\\
&&\!\!\!\!\!\!\times
\Big\vert\frac{c^2 p_x(\eta,\mf{q})}{\varepsilon_\mathbf{q}\sqrt{\omega\I{H}}}\langle0\vert\mf{p}(\eta,\mf{q})\rangle\Big\vert^2 \Big\vert \langle \mf{p}(\eta',\mf{q})\vert E\I{X} x\vert 0\rangle \Big\vert^2 \nonumber
\end{eqnarray}
for the considered relativistic setup [see also Eq.~\eqref{mhhg4a}], and by~\cite{MILOSEVIC:RL-02}
\begin{eqnarray}\label{eq:1atom_nonrel}
 \frac{\D w_n}{\D \Omega}{\Big|}\I{t,s}&=&\frac{\omega\I{L}^2}{(2\pi c)^3} \omega\I{H}^2 \Bigg|\frac{1}{\omega\I{L}^2} \sqrt{\frac{(-2\pi i)^5}{D(\mf{p},\eta,\eta')}}\Bigg|^2
\nonumber\\
&&\!\!\!\!\!\!\!\!\!\!\!\!\!\!\!\!\!\!\!\!\!\!\!\!\!\!\!\!\!\!\!\times\Big\vert\sqrt{\omega\I{H}}\langle0\vert x\vert p_x+A(\eta)/c\rangle\Big\vert^2 \\
&&\!\!\!\!\!\!\!\!\!\!\!\!\!\!\!\!\!\!\!\!\!\!\!\!\!\!\!\!\!\!\!\times\Big\vert \frac{2\sqrt{2}\kappa^3}{\vert E(\eta')\vert} \langle p_x+A(\eta')/c\vert V\vert 0\rangle e^{-\imag(\tilde{S}(\mathbf{q},\eta,\eta')+\frac{\omega\I{X}}{\omega\I{L}} \eta'-\frac{\omega\I{H}}{\omega\I{L}}\eta)}\Big\vert^2 \nonumber
\end{eqnarray}
for a conventional nonrelativistic setup where we inserted the tunneling correction factor of Ref.~\cite{MILOSEVIC:RT-02} and where $D(\mf{p},\eta,\eta')=\mathrm{det}\frac{\partial \tilde{S}(\mf{p},\eta,\eta')}{\partial (\mf{p},\eta,\eta')}$ is the functional determinant of the respective action.
Both,~Eq.~\eqref{eq:1atom_rel} and~Eq.~\eqref{eq:1atom_nonrel} are evaluated at the saddle point belonging to the short trajectory of the respective energy.
All factors in Eq.~\eqref{eq:1atom_rel} and Eq.~\eqref{eq:1atom_nonrel} were ordered in the same way and a distinct physical meaning can be assigned to them~\cite{IVANOV:CO-96,KOHLER:HX-11}:   
\begin{equation}
\frac{\D w_n}{\D \Omega}
\propto \omega\I{H}^2  \left(\vert \mf{p}_f\vert \frac{\partial\omega\I{H}}{\partial t\I{i}}\right)^{-1}
\left| a\I{rec}(\mf{p}\I{f})\right|^2\frac{\D^3 w\I{i}(t\I{i},\mf{p\I{i}})}{\D \mf{p\I{i}}^3} \, .
\label{wn}
\end{equation}
The factor $\omega\I{H}^2$ accounts for the phase space and converts the matrix element into the probability, $\D^3 w\I{i}(t\I{i},\mf{p\I{i}})/\D \mf{p\I{i}}^3$ is the differential ionization rate with the ionization time $t\I{i}$ in momentum direction $p\I{i}$, $a\I{rec}(\mf{p}\I{f})$ is the recombination amplitude and the last factor accounts for the dynamical properties of the wave, $\mf{p}\I{f}$ is the final momentum at recollision, $ \partial \omega\I{H}/\partial t\I{i}$ is the electron wave packet chirping factor discussed in~\cite{KOHLER:HX-11}.  

We compare all factors in~Eq.~\eqref{wn} between the two cases, to identify the reasons for the five orders of magnitude suppression of the single-atom yield in the relativistic regime. Since the harmonic energy increases by a factor of $10^3$ by going to relativistic case, the factor $\omega\I{H}^2$ yields an increase of $6$ orders of magnitude. The differential ionization rate of the particular trajectory is reduced by a factor of $10^{-2}$. Three properties contribute to estimate this ratio: The electron angular distribution of ionization is much broader for the one-photon ionization $p^2\sim(\omega\I{X}-I\I{p,x})\sim6$ than for tunneling ionization $p^2\sim\frac{3E}{\sqrt{2 I\I{p,t}}}\sim0.23$~\cite{HATSAGORTSYAN:UFO-08} yielding a factor of $4\times 10^{-2}$. On the other hand, the total (constant) ionization rate of the relativistic example [see Eq.~\eqref{eq:io-rate-zr}] is by a factor 2 higher than the instantaneous rate of non-relativistic example~\cite{PERELOMOV:IA-67,AMMOSOV:TR-86}. Third, in the relativistic setup, the relevant electron trajectory starts with a certain angle $\theta$ off the x-ray field direction resulting in a factor $\cos^2\theta\approx0.3$ in Eq.~\eqref{eq:io-rate-zr}. Together, we find the ratio of the differential rate $0.3\times2\times4\times 10^{-2}\approx 10^{-2}$. Note that the total time-average ionization rate is a factor of 10 lower for the non-relativistic example than for the relativistic one [see Eq.~\eqref{eq:io-rate-zr}]. The recombination amplitude is reduced by a factor of $10^{-5}$ as discussed in~\cite{KOHLER:HX-11}. The factors $p\I{f}$ and $\partial\omega\I{H}/\partial t\I{i}$ contained in the functional determinant reduce the relativistic yield by $10^{-4}$.
Collecting all factors, we find the suppression of $10^6\times10^{-2}\times10^{-5}\times10^{-4}=10^{-5}$ of the single-atom yield according to ratio between the respective terms in Eq.~\eqref{eq:estcprop} and~Eq.~\eqref{eq:estmpy}.

\section{Conclusion }
\label{sec:conclusion}
Extending table-top HHG to the hard x-ray domain is an exciting prospect, especially because many research labs already use HHG as a XUV sources and other approaches to generate hard x rays, like free electron lasers, require large scale facilities.

The present study discussed several difficulties that need to be overcome in order to realize the idea. The relativistic drift has been extensively discussed in the literature. Each proposed geometry has its own advantages and disadvantages regarding  phase-matching. 
Generally, increasing the harmonic energy renders phase-matching more difficult for many reasons: the emission phase of the harmonics depends approximately linearly on the intensity ($\phi\sim U\I{p}\tau$). Small intensity variation, e.g., in a Gaussian focus, immediately results in phase difference much larger than $\pi$. On the other hand, differences in the phase velocities between the harmonics and the laser wave lead to a slip in space between both waves. This results in phase-mismatch as soon as the slip is comparable to the harmonic wavelength which happens earlier for shorter harmonic wavelengths. Additionally, relativistic HHG is always accompanied by a large ionization 
leading to an enormous plasma dispersion.
For these reasons, in the best case, we obtain realizable medium lengths of only a few tens of~$\mu$m reducing the expectable macroscopic yield. 

Apart from the relativistic drift and phase-matching, we identified further issues decreasing the harmonic emission in relativistic HHG connected with the single-atom yield. First, recombination of the recolliding electron becomes less likely for high momenta: scattering is favored instead. Secondly, the electronic wave function is spread over a larger energy bandwidth. If phase-matching cannot be achieved for the whole bandwidth, however, a large part of the harmonic radiation is lost. This was expressed by the chirping factor. Third, the solid emission angle  decreases quadratically with the harmonic energy increase.

Regarding the harmonic yield, the present setup for relativistic HHG as well as the one in~\cite{KOHLER:HX-11} yield a small photon number for emitted harmonics that are both of similar order. On the bottom line, we think that the setup considered in this paper is more promising than that of~\cite{KOHLER:HX-11} because the required laser intensities are lower and the phase-matching scheme is more practical. 

One important conclusion of the paper is that phase-matching favors the collinear alignment of the laser and x-ray beams for the x-ray assisted relativistic HHG setup.
This co-propagation is sufficient to induce drift compensation and no perpendicular alignment of both beams is required. Note that the collinear geometry has already been used in various experiments~\cite{ISHIKAWA:PI-03,TAKAHASHI:DE-07,GAARDE:DE-05}.  

\section{Acknowledgements}

We would like to thank Christoph H. Keitel and Michael Klaiber for fruitful discussions.

\bibliography{kohlerref}

\end{document}